**Deep Representation Learning of Tissue Metabolome and Computed Tomography Images Annotates Non-invasive Classification and Prognosis Prediction of NSCLC**

*Marc Boubnovski Martell[1], Kristofer Linton-Reid[1], Sumeet Hindocha[2], Mitchell Chen[1], OCTAPUS-AI, Paula Moreno[3,5], Marina Álvarez-Benito[3,6], Ángel Salvatierra[3,6], Richard Lee[2,7], Joram M. Posma[1], Marco A Calzado[3,4]\* and Eric O Aboagye[1]\**

[1] Imperial College London Hammersmith Campus, London, SW7 2AZ, United Kingdom.
[2] Early Diagnosis and Detection Centre, National Institute for Health and Care Research Biomedical Research Centre at the Royal Marsden and Institute of Cancer Research, London, SW3 6JJ, United Kingdom.
[3] Instituto Maimónides de Investigación Biomédica de Córdoba (IMIBIC), Córdoba, 14004, Spain.
[4] Departamento de Biología Celular, Fisiología e Inmunología, Universidad de Córdoba, Córdoba, 14014, Spain.
[5] Departamento de Cirugía Toráxica y Trasplante de Pulmón, Hospital Universitario Reina Sofía, Córdoba, 14014, Spain.
[6] Unidad de Radiodiagnóstico y Cáncer de Mama, Hospital Universitario Reina Sofía, Córdoba, 14004, Spain.
[7] National Heart and Lung Institute, Imperial College London, Guy Scadding Building, Dovehouse Street, London, SW3 6LY, United Kingdom.

*Email: m.boubnovski-martell18@imperial.ac.uk, k.linton-reid18@imperial.ac.uk, sumeet.hindocha@rmh.nhs.uk, mitchell.chen@imperial.ac.uk, pmoreno39@gmail.com, marinaalvarezbenito@telefonica.net, asalvati@separ.es, richard.lee@rmh.nhs.uk, j.posma11@imperial.ac.uk, bq2cacam@uco.es, eric.aboagye@imperial.ac.uk*

**Abstract**
The rich chemical information from tissue metabolomics provides a powerful means to elaborate tissue physiology or tumor characteristics at cellular and tumor microenvironment levels. However, the process of obtaining such information requires invasive biopsies, is costly, and can delay clinical patient management. Conversely, computed tomography (CT) is a clinical standard of care but does not intuitively harbor histological or prognostic information. Furthermore, the ability to embed metabolome information into CT to subsequently use the learned representation for classification or prognosis has yet to be described. This study develops a deep learning-based framework -- tissue-metabolomic-radiomic-CT (TMR-CT) by combining 48 paired CT images and tumor/normal tissue metabolite intensities to generate ten image embeddings to infer metabolite-derived representation from CT alone.
In clinical NSCLC settings, we ascertain whether TMR-CT achieves state-of-the-art results in solving histology classification/prognosis tasks in an unseen international CT dataset of 742 patients. TMR-CT non-invasively determines histological classes - adenocarcinoma/ squamous cell carcinoma with an F1-score=0.78 and further asserts patients' prognosis with a c-index=0.72, surpassing the performance of radiomics models and clinical features. Additionally, our work shows the potential to generate informative biology-inspired CT-led features to explore connections between hard-to-obtain tissue metabolic profiles and routine lesion-derived image data.

# 1. Introduction

Numerous studies have developed imaging analysis pipelines to analyze and diagnose lung cancers. Tools such as radiomics have been helpful in extracting features from CT scans of lung cancer patients, followed by machine learning models to perform classification or prognosis tasks.[1,2,3] These features quantify the tumor's spatial complexity, such as shape, size and intensity features and are becoming part of a routine investigation in the literature.[3] More recently, some studies have attempted to replace radiomic features with deep learning (DL) features extracted from convolutional neural networks (CNN) directly on lesions.[4,5,6,7] Of various CNN architectures, autoencoders are some of the most widely adopted and aim to find features that allow a model to reconstruct the original image in a different context.[4,7,8] However, in practice, these features have limited clinical performance, e.g. classification C-index = 0.65, and the used framework u precludes intuitive biological or clinical interpretability.[9] Current studies generate radiomic features and subsequently check if they associate with genes, metabolites, proteins and other biological factors, using for example, gene-set enrichment analysis.[10,11,12] In contrast, our present study develops a framework for generating features from images that have already learned specific biological representations of tissue metabolites.

A new and evolving field of computational biology combines two or more diverse modalities to improve the performance of each.[13,14,15] Gundersen and colleagues demonstrate the feasibility of developing multimodal pairings of pathology and genomic profiles to extract deep features from pathology images connected to the genomic profiles of the patient to obtain more explainability.[16] The corollary of this learned representation approach, once developed, implies that one of the two modalities is sufficient to represent the other in the absence of both modalities being present.[16] Inspired by this approach, we investigate whether it is possible to generate deep features from hard-to-obtain tumor and normal tissue metabolome data, on the one hand, and the more routine CT scan image data on the other.

In characterizing tumors, the choice of metabolomics profiles as a benchmark is predicated on our recent work that expounds the use of tumor and adjacent tissue metabolome information in asserting the classification of histology subtypes, achieving an F1-score of 0.96, significantly outperforming most published models from imaging data in the field of lung cancer subtype classification or prognosis.[17,18] While the chemical information from tissue metabolomics is rich, the approach is not routinely used in patient management due to its invasiveness and analytical complexity.[19] Thus, we developed a pipeline using an autoencoder to investigate, for the first time, whether deep features of CT image reconstructions linked to chemical information from the metabolomics of patients will provide sensitive clinical information. These deep features extracted from the deep probabilistic canonical correlation analysis (DPCCA) model were named tissue metabolomic radiomic computed tomography (TMR-CT) as the model aims to connect both data modalities.

Specifically, the model is composed of a two-stage neural network and PCCA; the neural network part first finds embeddings for each data modality (CT scan, metabolites) separately and then combines these by maximizing the correlation whilst minimizing the reconstruction loss of both modalities.[16] The benefit of this structure is two-fold. First, the deep features captured for each data view maximize the shared variation. Second, the generative structure of the model allows cross-data modality imputation. This is particularly important given the difficulty in obtaining the paired datasets from both modalities. [17] We explore the aforementioned benefit in this manuscript by utilizing the embeddings derived from CT scans for histology subtype classification and prognosis of non-small cell lung cancer (NSCLC) patient data. Our approach achieves state-of-the-art results in both tasks while also providing valuable biological insights. We accomplish this by employing TMR-CT, which encompasses reconstructed representations of tissue metabolite types and intensities. This work demonstrates the potential for the method to enhance the practitioner's – radiologist's, respiratory physician's, or oncologist's – ability to determine histology subtype classification, as well as prognosis, using algorithms derived from the current work, as illustrated in **Figure 1**.

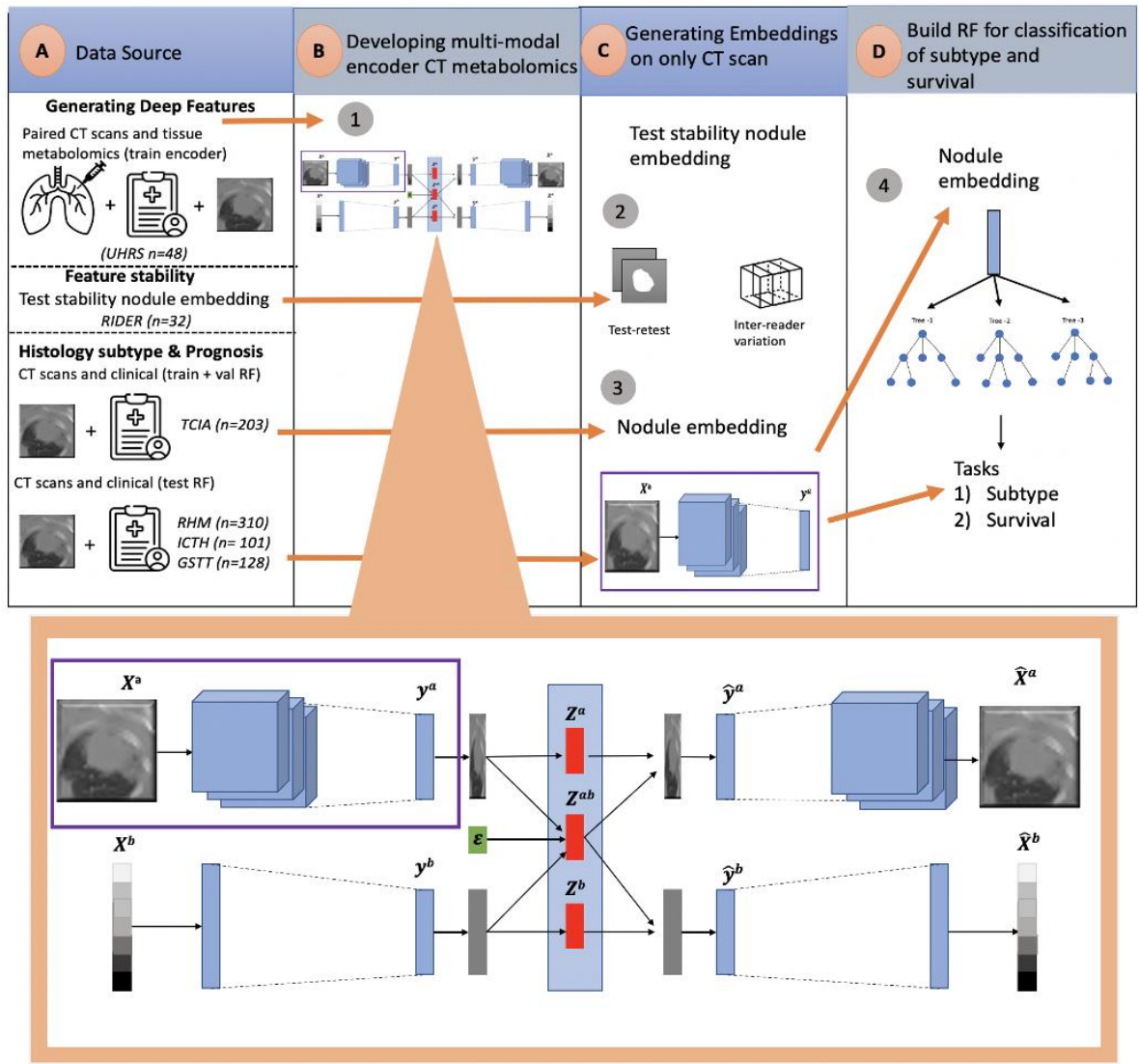

**Figure 1.** The workflow of our study is separated into four sections: (A) Dataset collection for: generating deep features, evaluating feature stability, histology subtype classification and prognosis. (B) The DPCCA model is used to find a shared latent space between the CT scans and metabolomics; an enlarged version of this model is shown – the part in the purple box is the section used to generate TMR-CT features. In this model, $X^a$, $X^b$ is the original paired image and metabolomics; from these, we create $y^a, y^b$ image and metabolomics embeddings, respectively. The PCCA model then combines them into $z^a, z^{ab}, z^b$ latent variables. The latent variables $z^a, z^b$ capture view-specific variation whilst $z^{ab}$ captures the covariance. From the latent variables, we use a generative process of the model-sampling from the low dimension PCCA to reconstruct image and metabolomics embeddings $\hat{y}^a, \hat{y}^b$. Each embedding $\hat{y}^a, \hat{y}^b$ is then decoded to produce the $\hat{X}^a, \hat{X}^b$ using view-specific decoders. During implementation, when we only have the CT image, we extract the learned representation $y^a$, which we have defined as the TMR-CT features. (C) Convolutional neural networks in the purple box was used to generate TMR-CT features on external datasets, as well as test the stability of the features using the RIDER dataset. (D)

Use of TMR-CT features for histology subtype classification; we show random forest (RF)-based approaches as those were the ones that performed the best on both tasks.

## 2. Results

### 2.1 Datasets

The datasets in our study can be split into three parts, as seen in Figure 1 (data source): developing deep features, testing feature stability, histology classification and prognosis.

The dataset from 48 patients with both tissue and CT scans used to develop TMR-CT was obtained from the University Hospital Reina Sofia (UHRS), Spain. The research study was conducted in accordance with the Helsinki Declaration and was approved by the Cordoba Clinical Research Ethics Committee, all patients provided a signed written informed consent for participation in the study. Paired CT and tissue, obtained from both the tumor and non-tumor adjacent tissues, were collected from patients with NSCLC. Tissue samples were stored by the Andalusian Health Services Biobank, and the metabolomic profiling was performed under contract by Metabolon.[17] The patients did not receive any radiation or chemotherapy treatments before surgical resection, and the clinicopathological information was obtained prospectively and shown in **Table 1**. All tissue data were processed as previously reported.[17] CT scans were segmented by a board-certified clinical radiologist (MC).

Metabolite analyses were performed as previously described.[17] Samples were extracted by an aqueous methanol extraction process and analyzed with ultra-performance liquid chromatography/tandem mass spectrometry (UPLC/MS/MS; positive mode), UPLC/MS/MS (negative mode), and GC/MS by Metabolon. Tissue metabolites were identified by comparison with library entries of purified standards or recurrent unknown entities. Based on the literature and KEGG/HMDB databases, metabolites were annotated to one of eight 'super pathways' corresponding to their general metabolic processes (amino acid, lipid, carbohydrate, nucleotide, peptide, energy, cofactors and vitamins, and xenobiotics), and to one of 73 'sub pathways' representing more specific metabolic pathways or biochemical subclasses; in the aggregate, 851 metabolites were identified through this approach for both lung adenocarcinoma (AC) and squamous cell carcinoma (SCC) subtypes, and normal lung tissues.[17]

**Table 1.** Patient demographics in the dataset from University Hospital Reina Sofia, Spain, with joint CT and metabolomics used for developing TMR-CT.

|  | Characteristics | (n=48) | |
|---|---|---|---|
|  |  | AC[†] | SCC[†] |
|  | Patients | 22 | 26 |
|  | Age | 61.7(±16.9) | 70.3(±7.3) |
| **Gender** | Male | 16 | 26 |
|  | Female | 6 | 0 |
| **TNM8 Overall stage** | 1 | 18 | 17 |
|  | 2 | 3 | 7 |
|  | 3 | 1 | 2 |

[†]AC, adenocarcinoma of the lung; SCC, squamous cell carcinoma of the lung.

To test the stability of the TMR-CT features, we used the open-sourced RIDER dataset consisting of 32 patients with NSCLC who underwent two sequential chest CT scans within 15 mins, employing the same imaging protocols. [20] In this study, three radiologists measured the two greatest diameters of each lesion on both scans obtaining highly reproducible measurements, all with concordance correlation coefficients (CCC) greater than 0.96. Thus, this dataset has been shown to be useful in determining the reproducibility of deep learning features for NSCLC. [21]

To test how useful TMR-CT is for histology classification and prognosis prediction, we used four different datasets summarized in **Table 2**. To train our models, we used the open-source TCIA (The Cancer Imaging Archive), from which we selected 203 patients diagnosed with either AC or SCC.[22] The TCIA was split into 120 for training and validation and 83 for external validation (ext val).Then, to evaluate how well the developed model generalize to new NSCLC datasets, we used three geographically distinct datasets from the OCTAPUS-AI study (ClinicalTrials.gov identifier: NCT04721444) as external test sets (GSTT, Imperial and RMH); OCTAPUS-AI represents a study from multiple UK cancer centers (Guy's and St Thomas' NHS Foundation Trust, Imperial College Healthcare NHS Trust and the Royal Marsden NHS Foundation Trust respectively) collected for the explicit purpose of developing robust predictive lung cancer algorithms.[23]

**Table 2.** Patient demographics and treatment variables in the three external datasets used for histology classification and prognosis (for age and radiotherapy dosage, we show the median value together with the IQR in brackets).

| | | Train and Validation | | External test | | | | | |
|---|---|---|---|---|---|---|---|---|---|
| | | TCIA (n=203) | | GSTT (n=128) | | Imperial (n=101) | | RMH (n=310) | |
| | Characteristic | AC | SCC | AC | SCC | AC | SCC | AC | SCC |
| | Patients | 152 | 51 | 67 | 61 | 49 | 52 | 189 | 121 |
| | Age | 68(±15) | 71(±14) | 70(±15) | 73(±11) | 71(±14) | 72(±11) | 74(±17) | 76(±12) |
| Gender | Male | 32 | 112 | 36 | 39 | 27 | 33 | 83 | 82 |
| | Female | 19 | 40 | 31 | 22 | 22 | 19 | 106 | 39 |
| CT type | Contrast | | | 29 | 23 | 29 | 31 | 55 | 42 |
| | non-contrast | | | 38 | 37 | 20 | 21 | 134 | 79 |
| Dosage | Biologically effective dosage, Gy | | | 77(±39) | 77(±35) | 70(±9) | 70(±9) | 77(±39) | 72(±23) |
| End result | Survival days | 583.0 | 492.0 | 864.0 | 760.0 | 895 | 867.7 | 834.0 | 694.0 |
| | Recorded deaths | 45 | 139 | 32 | 40 | 26 | 42 | 104 | 79 |
| Treatment | Conventional RT only | | | 10 | 19 | 22 | 29 | 31 | 31 |
| | SBRT | | | 29 | 17 | 0 | 0 | 90 | 31 |
| | Sequential chemoRT | | | 10 | 11 | 9 | 9 | 37 | 37 |
| | Concurrent chemoRT | | | 18 | 14 | 18 | 14 | 31 | 18 |
| TNM8 Overall Stage | 1 | | | 32 | 23 | 10 | 8 | 89 | 36 |
| | 2 | | | 10 | 9 | 13 | 12 | 22 | 22 |
| | 3 | | | 25 | 29 | 26 | 32 | 78 | 63 |
| | 2 | | | 0 | 0 | 0 | 0 | 172 | 114 |

| | | | | | | | |
|---|---|---|---|---|---|---|---|
| Slice thickness | 2.5 | 67 | 61 | 0 | 0 | 17 | 7 |
| | 3 | 0 | 0 | 49 | 52 | 0 | 0 |

## 2.2 Overview of Metabolic Profiles for NSCLC

With many more metabolomic features than patients, we first filtered the metabolomics by only including those profiled in all 48 patients for both tumor and non-tumor adjacent tissue from the UHRS hospital; this reduced the number of metabolites to 174. The super pathway of these features is summarized in **Figure 2a**, and we observed a high degree of positive and negative correlation between several of the features from the tour tissue samples, as shown in **Figure 2b**.

Despite the large number of metabolomic features compared to our sample size, we did not need to perform feature reduction, as principal component analysis (PCA) is known to be robust to correlated features.[24] Furthermore, we reasoned that when incorporating the metabolomics feature into the DPCCA model, we would perform data augmentation as specified in section 4.2 to mitigate overfitting.

We were unable to use permutation importance to identify the most important metabolomic features due to feature collinearity; permuting any single feature would have little effect on the random forest (RF) performance. As an alternative, we performed hierarchical clustering on the Pearson rank-order correlation and chose a single feature from every cluster, as suggested by Rosato and co-workers in a systems biology-enhanced analytical framework for metabolomics data.[25] This approach allowed us to reduce the number of features to 6 metabolomics (1,5-anhydroglutocitol (1,5-AG), 1-arachidonoylglycerophosphoethanol-amine*, 1-stearoylcerol (1-monostearin), 3-hydroxybutyrate(BHBA), 3-phosphogylcerate and alanine) while still maintaining the same performance (F1-score of 1) in discriminating AC from SCC tissue.

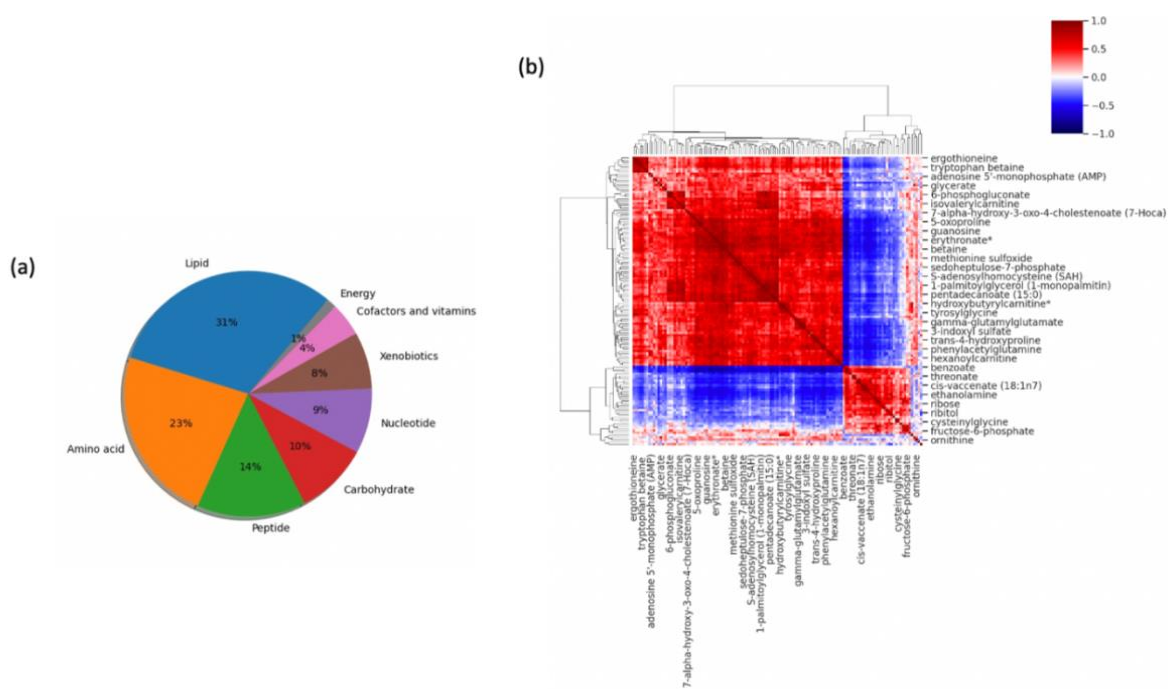

**Figure 2.** Metabolomic information on UHRS (a) Distribution of metabolites super pathway, which were present in all patients data used in the current study (b) Pearson correlation heat map between tumor metabolomics for all patients.

## 2.3 Ordinarily Metabolites discriminate histology subtypes but are unconnected to radiomic features

We investigated the data structure of the metabolomic profiles obtained from 48 tumor and non-tumor tissue samples in relation to the radiomic features from CT scans. For each CT scan, we extracted 665 radiomic features using the TextLab 2.0 software related to shape, size, intensity, and wavelet decomposition.[1] After pairing metabolomic and radiomic features for the 48 samples, we compared the predictive power and connection between both modalities. To determine the predictive power of both modalities, we examined 2-dimensional Principal Component Analysis (PCA) with all variables. We found that metabolomics provided was more informative when predicting tissue subtypes compared to CT radiomics **Figure 3(a-c).** Determining the most important metabolomic features, proved to be challenging as described in section 2.2 and we used hierarchical clustering based on the Pearson rank-order correlation.

Due to the large number of radiomic features, it is common practice to perform a feature reduction step prior to model building. While there is no established set way, various task-dependent strategies have been proposed.[26] For comparison, we only retained features with an intra-class-correlation coefficient (ICC) greater than 0.75, resulting in 438 features.[27]

To investigate the connection between the two sets of features within the tumor metabolomics and imaging datasets, we conducted a Pearson correlation analysis between the radiomics and the top ten metabolomics features. A heatmap of the data is shown in **Figure 3(d)**With a maximum absolute correlation coefficient of 0.45 and a mean correlation of 0.02, the results suggested a weak correlation between the two data modalities and techniques. Consequently, traditional canonical correlation analysis might be inappropriate.

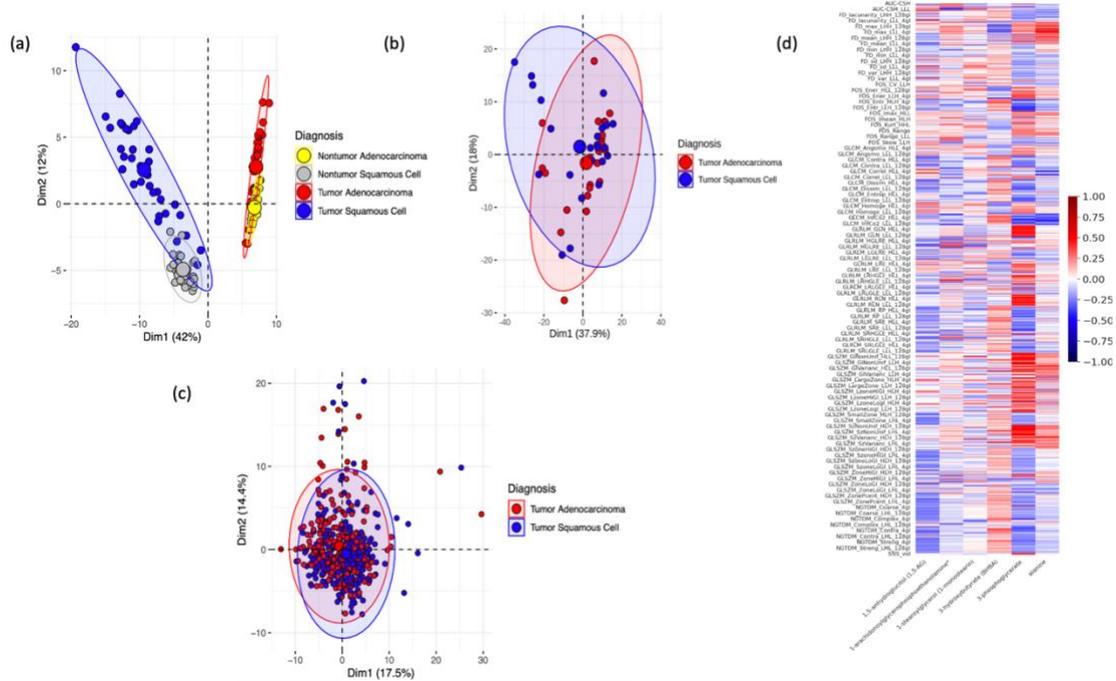

**Figure 3.** Comparison of Metabolomic and Radiomic features for AC and SCC classification of NSCLC patients (a) Two-dimensional PCA of metabolomic features from UHRS dataset (b) Two-dimensional PCA of radiomic features from UHRS dataset (c) Two-dimensional PCA of radiomic features from a larger TCIA dataset (d) Pearson correlation heatmap between CT radiomic features and six metabolomic features important for classification of histology subtypes from UHRS dataset.

## 2.4 The DPCCA model can be trained on CT scans and metabolomics to define TMR-CT

DPCCA was trained to expound the shared latent space between metabolomics and CT scan data. To evaluate the performance of DPCCA, we assessed the following. Firstly, we investigated if our model could reconstruct both modalities from the shared latent space. Thus, we examined the reconstructed CT slices and the metabolomic covariance matrix on the held-out test dataset. As seen in **Figure 4**, DPCCA successfully reconstructed both views. As enshrined in a similar inference framework by Gundersen and co-workers for gene expression and pathology latent space,[16] we wanted to verify that our end-to-end model, composed of a neural network and DPCCA, makes use of both components. To test this, we computed the expected complete negative log-likelihood on the held-out dataset. We showed that the loss function decreased during training, supporting the notion that the embedding model uses both deep learning and neural networks to create embeddings and reconstruct the data from nonlinear observations.

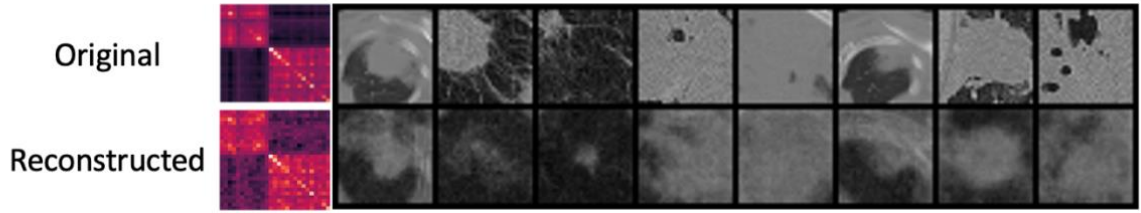

**Figure 4.** Reconstructing the metabolite-inspired CT scan. To test the quality of our latent space model developed by DPCCA, we examined CT and metabolomics reconstruction. The images above were obtained from the test data **for different patients.** (Top row) The original metabolomics expression covariance matrix and random CT slices from test data and (Bottom row) the reconstruction of the CT image of unseen test dataset when both the original image and metabolites are provided as inputs to the model.

To interpret the influence of metabolomics on the shared embeddings, we plotted the correlations. **Figure 5(a)** shows the presence of clusters with relevant metabolite super pathway information on the test dataset of the UHRS. To gain a good understanding of metabolomic features that are the focus of the study, we plotted the maximum absolute correlation between the metabolomic features and the ten TMR-CT in **Figure 5(b)**. The two metabolites with the lowest correlation are 2-hydroxyglutarate and urea, with a maximum correlation of 0.14 and 0.16, respectively. The two metabolites with the highest correlation are sedoheptulose-7-phosphate and uridine, with a correlation of 0.72 and 0.64, respectively. It is important to note that the focus of the DPCCA model was to find a shared latent space and then reconstruct the learned representation; thus the 'metabolomic features' that the model focuses on are not necessarily those with the greatest classification or prognosis power, but rather those that the model can use in representing the CT-images.

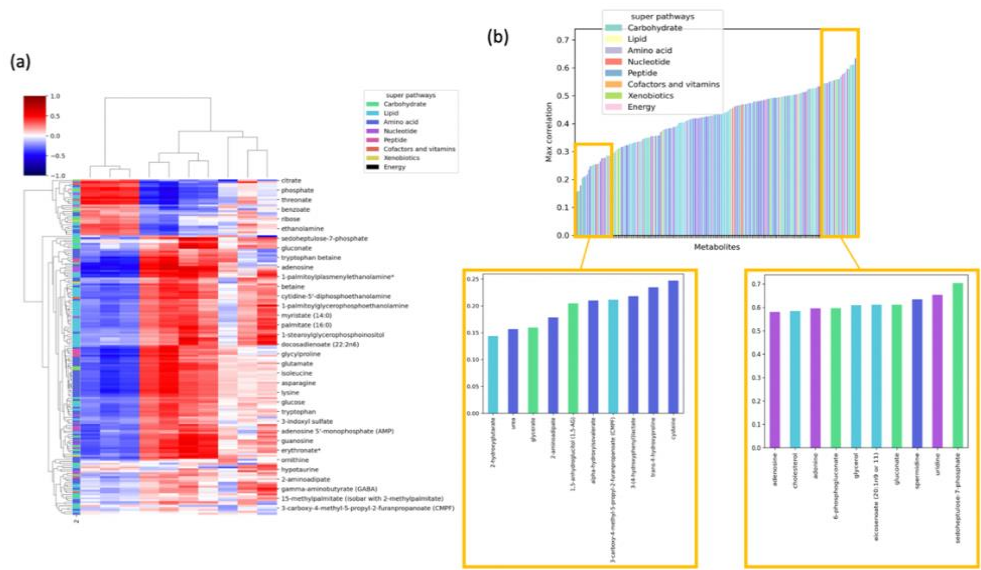

**Figure 5.** (a) Pearson correlation heatmap between tumor metabolomic features and TMR-CT features from DPCCA in figure 1 (b) Correlation of performance of the TMR-CT model developed by DPCCA, the bar plot shows the highest absolute correlation between metabolites and the TMR-CT features; we expand the top ten least correlated metabolites and the top ten most correlated metabolites in the yellow boxes.

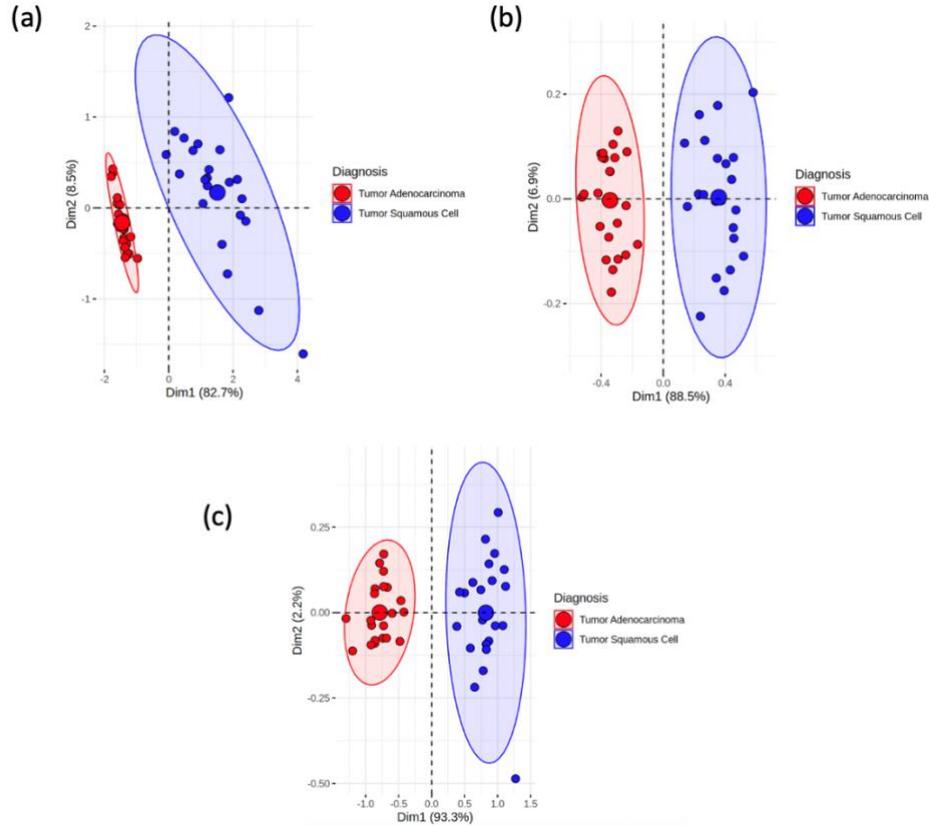

**Figure 6.** PCA applied on the UHRS applied to the three different features (a) TMR-CT (image embeddings space $y^a$) (b) metabolomic embeddings $y^b$ and (c) image and metabolomic shared latent space $z^{ab}$.

## 2.5 Reliability and reproducibility of TMR-CT

To assess the stability of our encoder algorithm, we tested it in a test-retest context using the publicly available dataset, RIDER, consisting of 32 patients with lung cancer. Each patient underwent two chest CT thorax scans (within 15 min apart) using the same imaging protocol. [20] We evaluated the stability of the encoder by examining the TMR-CT features between the test and retest scans. Our results demonstrated a high level of stability with an ICC of 0.86 for TMR-CT showing that our model had been well-regularized.

To account for inter-reader stability, we adopted an approach of relocating the input seed points to the center of the tumor. This aimed to simulate various radiologists annotating the tumor, which would cause variability between them. In this case, we showed a high correlation with a Spearman's rank-order correlation of 0.85 between the TMR-CT, showing strong inter-reader stability.

## 2.6 Exploiting TMR-CT features from DPCCA for classification and prognosis of CT scans without metabolomic profiles

We aim to determine if our latent variables captured meaningful, held-out biological information such as histology subtype and overall survival (OS). The models were trained and the hyperparameters were tuned on TCIA data as seen in **Figure 7**, to select the best model for each task. In the case of the radiomics features we also selected the best feature selection technique.

From **Figure 7**, it is evident how RF using TMR-CT features significantly outperforms the traditional radiomics features extracted using TextLab 2.0 for histology classification and prognosis without performing any feature selection. This finding is important as it shows that the quality of the TMR-CT features is sufficiently high. Thus, no feature selection is required. This can be further seen in **Tables 3 and 4** where we chose the best performing feature selection and model, respectively, from Figure 7 to report the results on the four external datasets.

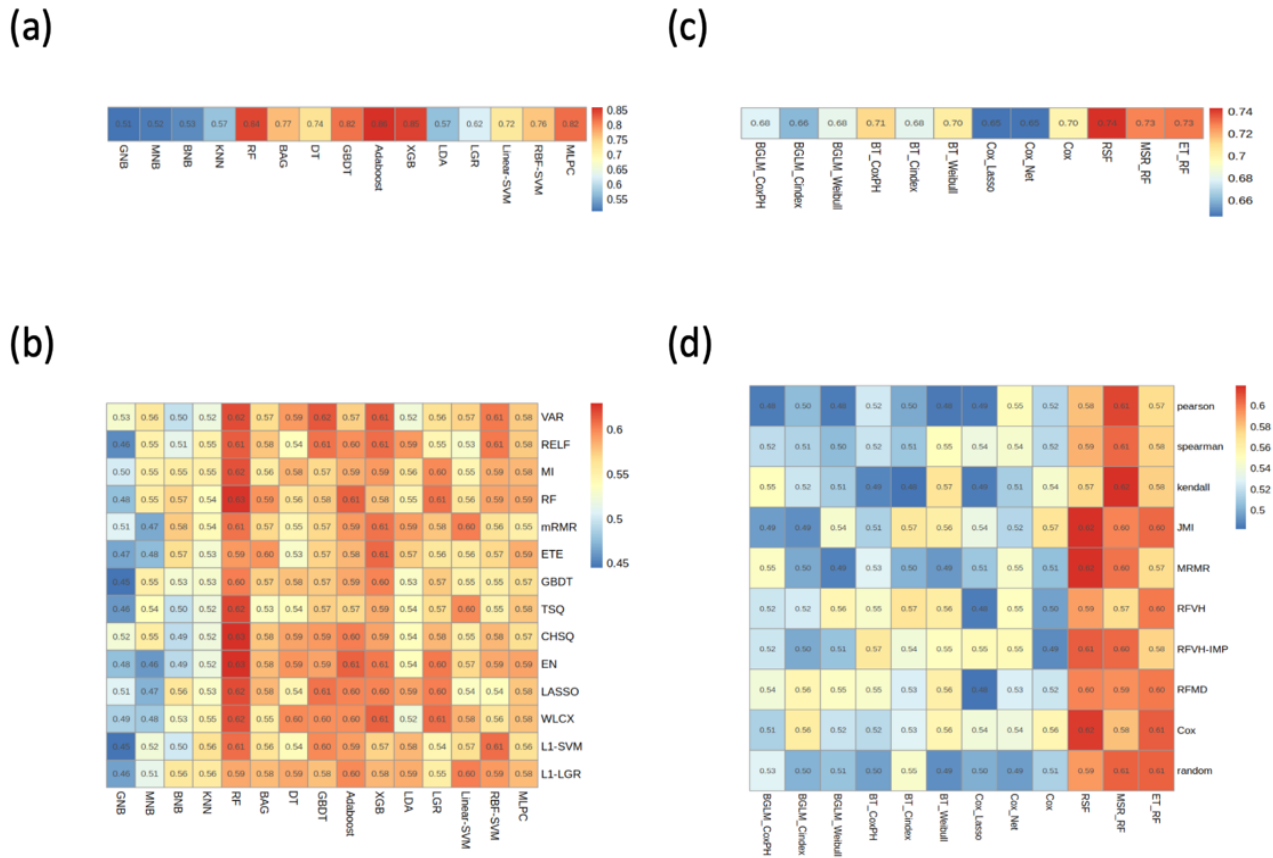

**Figure 7.** Classification and prediction heatmaps. (a) F1-score for classification of AC and SCC using TMR-CT for different classification models. (b) F1-score for classification of AC and SCC using radiomic features, with the x-axis being the predictive models and the y-axis corresponding to feature selection techniques (c) C-index for a prognosis for different prognosis models using TMR-CT. (d) C-index using radiomic features, with the x-axis being the prognosis models and the y-axis corresponding to feature selection techniques.

**Table 3.** F1-score of RF for classification of AC and SCC reported using the best feature selection and machine learning model ± standard error.

|  | F1-score | | | |
| --- | --- | --- | --- | --- |
|  | TCIA external validation (n=84) | RMH (n=320) | GSTT (n=128) | ICHT (101) |
| **Radiomics** | 0.63 ± 0.02 | 0.58 ± 0.04 | 0.59 ± 0.03 | 0.57 ± 0.02 |
| **TMR-CT** | 0.84 ± 0.03 | 0.78 ± 0.02 | 0.77 ± 0.03 | 0.79 ± 0.03 |

TCIA, RMH, GSTT and ICHT are The Cancer Imaging Archive, Royal Marsden Hospital (UK), Guy's and St Thomas' Hospital (UK) and Imperial College Healthcare Trust (UK), respectively.

**Table 4.** C-index of Random Survival Forest for prognosis of NSCLC reported using the best feature selection and machine learning model ± standard error. The prognosis was determined using the models alone or in a multivariable model with clinical features.

|  | C-index | | | |
| --- | --- | --- | --- | --- |
|  | TCIA external validation (n=84) | RMH (n=320) | GSTT (n=128) | ICHT (n=101) |
| **Radiomics** | 0.62 ± 0.04 | 0.58 ± 0.05 | 0.61 ± 0.04 | 0.59 ± 0.06 |
| **TMR-CT** | 0.74 ± 0.03 | 0.72 ± 0.04 | 0.71 ± 0.05 | 0.71 ± 0.04 |
| **Radiomics + clinical** | 0.64 ± 0.05 | 0.59 ± 0.06 | 0.62 ± 0.06 | 0.58 ± 0.07 |
| **TMR-CT + clinical** | 0.78 ± 0.06 | 0.73 ± 0.06 | 0.71± 0.04 | 0.71± 0.05 |

TCIA, RMH, GSTT and ICHT are The Cancer Imaging Archive, Royal Marsden Hospital (UK), Guy's and St Thomas' Hospital (UK) and Imperial College Healthcare Trust (UK), respectively.

The three external test sets of Kaplan Meier curves are shown in **Figure 8** and demonstrate good separation between high and low-risk groups with log-rank tests confirming a statistically significant difference a 5% level in the GSTT and 1% for the Imperial and RMH.

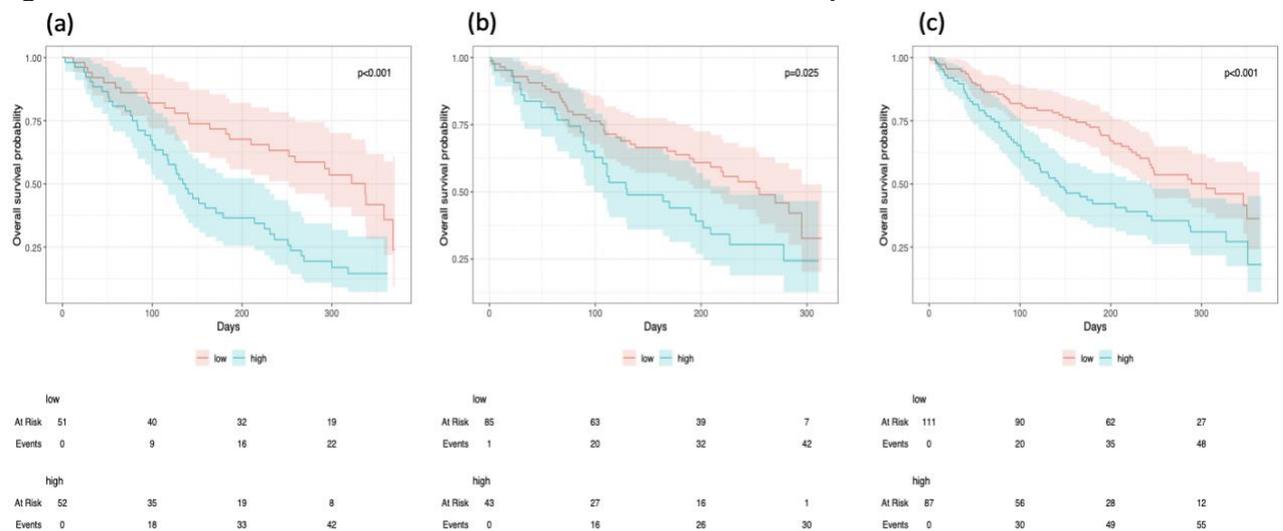

**Figure 8.** Kaplan Meir curves for low and high-risk groups of the dichotomised predicted probabilities using k-means clustering of the RSF with TMR-CT on the external validation: (a) Imperial, (b) GSTT and (c) RMH (*P*-values are from log-rank tests. Plots demonstrate good separation between high and low risk groups with log-rank tests confirming the statistical significance of 5% in the GSTT and 1% in Imperial and RMH.

To understand the importance of different features in our best prognosis model noted in Table 4, we reported the hazard ratio that our model calculated together with the P-value for the log-rank test for each feature modality in **Table 5**. We observe in a multivariable model that TMR-CT exhibited the highest hazard values in both the validation and external test datasets showing the high importance of TMR-CT for prognosis. This finding is notably more significant than clinical features, including age, gender, N-stage and gross tumor volume (GTV).

**Table 5.** Hazard Ratio (HR) and P-values for Permutation-Variable Importance Random Forest – Random Survival Forest (PVIRF-RSF) Model combining Clinical and TMR-CT ± standard error.

| | TCIA ext val | | GSTT | | Imperial | | RMH | |
|---|---|---|---|---|---|---|---|---|
| **Features** | HR(95%CI) | P-value | HR(95%CI) | P-value | HR(95%CI) | P-value | HR(95%CI) | P-value |
| **TMR-CT** | 38.0(30.3-45.7) | >0.00001 | 17.1(8.7-25.5) | 0.04 | 15.9(10.2-23.7) | >0.01 | 24.4(17.3-31.5) | >0.001 |
| **Age** | 1.24(0.83-1.65) | 0.3 | 2.14(1.63-2.65) | 0.73 | 1.04(0.66-1.41) | 0.84 | 1.15(0.92-1.38) | 0.34 |
| **Gender** | 1.53(1.12-1.94) | 0.2 | 3.20(2.91-3.49) | 0.06 | 1.05(0.58-1.53) | 0.92 | 3.02(2.21-3.83) | 0.03 |
| **N-Stage** | - | - | 1.16(0.89-1.43) | 0.12 | 2.33(1.76-2.90) | 0.03 | 4.30(2.93-5.67) | 0.12 |
| **GTV** | - | - | 1.04(0.52-1.56) | 0.65 | 1.00(0.46-1.64) | 0.5 | 1.15(0.71-1.59) | 1.14 |

## 3. Discussion

### 3.1 Summary of Main Results

In this study, we have shown that a deep learning framework - DPCCA - can model a connection between CT scans of lung nodules and their tissue metabolomics profiles against the premise that certain metabolites and/or their intensities, representing tumor growth and/or tumor microenvironment factors maximally co-vary together, linearly or non-linearly, with CT image features. Furthermore, we have shown the usefulness of such models, embodied within TMR-CT, for histology subtype classification and prognosis of NSCLC patients non-invasively, thus asserting clinical relevance. Notably, the DPCCA-generated learned representations can be used for downstream classification or prognosis tasks even when we only have CT scans available. Such metabolomic profile-correlated features are more interpretable biologically. This methodology would be useful in guiding treatment decisions, particularly in the context of patients that are unfit for biopsy.

The generated metabolite pairs are not inherently intuitive. For example, the most important metabolites (in our DPCCA model) differ from our previously published top metabolites from metabolomics-only analysis, in the study by Moreno et al. (2018). One of the main reasons for this is that, in the metabolomics-only study, the top metabolites are chosen as those most discriminative between tumor and non-tumor cases for AC and SCC separately, obtaining 20 different metabolites.[17] In the current study, however, the top ten metabolites chosen are those most correlated with our TMR-CT, whose purpose is to reconstruct the two data modalities as a composite phenotype of both the CT features of the lesion and the metabolomics profile. Regardless, the most correlated metabolites appear to regulate cell growth and membrane activity through glycolysis, pentose-phosphate, DNA synthesis and fatty acid metabolism.

Comparing the TMR-CT generated by DPPCA to radiomic features, we see that the machine learning algorithm using TMR-CT performs significantly better when performing two clinically distinct tasks histology classification and prognosis -- suggesting that TMR-CT derived features contain more information than the radiomic features even though there are significantly fewer features. This difference in performance was seen regardless of whether TMR-CT was used alone or combined clinically, meaning that TMR-CT has overall better performance than radiomic features for NSCLC histology classification and prognosis determination.

**3.2 Comparison to Previous Studies**
To our knowledge, over 16 different models have attempted to integrate multiomics using deep learning to gain a better understanding of the complex biological process of cancer.[15] However, most of these models aim to fuse multiomics of the same modality, making it a significantly easier process relying on the availability of both data modalities during the test time. Though some studies look at integrating imaging and tabular data, they only work when both modalities are present. [28] Thus, the benefit of using DPCCA over other models is that it can be applied during test time even without information about the tissue metabolomic profiles (with only the CT available). We show that TMR-CT derived from DPCCA was superior to conventional radiomics for histology classification and prognosis in patients who only had the CT scan available.

**3.3 Study Limitations**
Our current study exhibits several limitations. First, we only validated the performance of the TMR-CT on external validation datasets. Future analysis on a separate cohort with paired CT and metabolomics data would be required to validate the stability of the correlations identified in our study. Second, a wider range of histology could have been used. In our study, we only examined patients with AC and SCC, but our technique could easily be extended to incorporate other lung cancer histology with minimal adaptation.

**3.4 Possible Future Directions**

In the future, two primary directions could be explored by researchers. The first is to validate the TMR-CT features on an external dataset of paired CT scans and metabolomics features. The second is to increase the number of patients in the paired CT and metabolomics dataset to contain a larger number of patients that are more representative of the wider population by including small cell lung cancer patients. Unfortunately, such datasets don't currently exist, so we could not incorporate these ideas into our study. Nonetheless, by showcasing the efficacy of a niche algorithm in a specific context, we establish a foundation for future studies that aim to extend its performance and validation to diverse settings. By conducting a prospective study that combines TMR-CT, radiomics, and body fluid metabolomic analysis, it may be possible to improve prognostic capabilities when tissue metabolomics is unavailable. This is particularly relevant for patients who are deemed unsuitable for surgery or face obstacles in accessing tumor material for histology classification and prognostication prior to making a decision about surgery.

**3.5 Importance of Work**

Our study investigates the feasibility of using deep learning to combine patients' paired CT and steady-state metabolomics information to find a shared representation that can allow the reconstruction of both modalities. One benefit of using our two-step deep learning model is the ability to independently extract deep features from a single modality without needing another modality.

This is of specific importance in the clinical setting, where it is often the case that a single data modality is more readily accessible than the other. Enhancing the features obtained from lesions on CT images, we are improving the usefulness of CT scans, which are more readily available when evaluating NSCLC patients for early diagnosis and tumor prognostication. [29]

In summary, we were able to show that there is a connection between the metabolomic and CT features of NSCLCs. Furthermore, it is possible to exploit the learned representation within CT images of patients with NSCLC that co-vary with tissue metabolomic profiles and demonstrate their usefulness clinically for histology subtype classification and prognosis on external datasets when only a CT scan is present.

## 4. Experimental Section

### 4.1 Pre-processing of CT images

All image pre-processing was done using TorchIO, a package allowing for effective pre-processing of CT images. [30] To ensure comparability, the CT scans from all datasets were resampled to isotropic voxels of 1x1x1mm. This was performed using linear and nearest neighbor interpolation for the image and segmentation, respectively. [31]

### 4.2 Pre-processing data for training and testing DPCCA

We had 48 paired samples from the UHRS, each with two data views: CT scans and the metabolomic profiles of tumor/normal tissue. We performed a stratified split of the dataset into 32 paired samples for training, validation and 16 paired samples for testing whilst keeping the balance of AC/SCC consistent across splits. The following steps were done on the training and testing split separately.

Given the 3D segmentations, we calculated the center of mass (COM) and bounding box of the tumor. A 3D isotropic patch of 50x50x50, around the COM of tumor volume, was extracted, resulting in 48 3D tumor patches. We then created 3D patches of 32x32x32 randomly and ensuring that at least 65% of tumor was captured by the bounding box. The 3D patches were normalized to a range of 0-1 and lower upper boing of -1024 and 3021.[31]

For the metabolomics features, we only included those that were profiled in all patients for both tumor and non-tumor adjacent tissue. The reason for this was to increase the reproducibility of the chosen features. Subsequently, we normalized the values of the metabolomic features to have a mean of 0 and a standard deviation of 1.

Data augmentation makes it possible to increase the data available for training without actually collecting new samples by applying a range of techniques. In this study, an augmentation factor of 186,624 was applied to the patches resulting in a training dataset of approximately nine million 3D patches. These augmentations were chosen based on other similar studies and consisted of ±18 pixels in three axes, random rotations at 90° intervals along the longitudinal axes, and random flipping along three axes.[31] The augmentations were applied in real-time during training, and simultaneously, we applied Gaussian noise with a standard deviation of 0.1 to the image patches and metabolomic features. [31] No augmentation was applied during validation or testing.

### 4.3 Building DPCCA model

The DPCCA model is a deep generative model that fits the probabilistic canonical correlation analysis (PCCA) into two autoencoders, one for the CT image and the other for metabolomic. Figure 1 shows a detailed image of this model and where it fits the PCCA to the embeddings of two autoencoders. The code for this model was adapted from https://github.com/gwgundersen/dpcca. Specifically, we optimized the image autoencoder to enable studies with 3D images instead of 2D and used the 3D Deep Convolution Generative Adversarial Network (3D-DCGAN) developed specifically for medical images.[32] The model was trained end-to-end using the mean squared error (MSE) for regression model fitting of paired CT image and metabolomics data; and also, for the reconstruction of the loss function for the modalities separately. The following section details the DPCCA method and its adaptation to our task.

Given a paired sample ($\mathbf{x}^a$, $\mathbf{x}^b$), the linear and convolution encoder embedded the CT images and metabolomics, respectively. These embedded vectors $\mathbf{y}^a$ and $\mathbf{y}^b$ are then fitted by the PCCA and incorporate an $l_1$ penalty on the PCCA metabolomic weights, thus, encouraging sparsity in the metabolomic profiles and resulting in shared and view-specific latent variables $\mathbf{z} = [\mathbf{z}^{ab}\ \mathbf{z}^a\ \mathbf{z}^b]^T$.

Mathematically the PCCA can be expressed as follows by **Equation 1**:

$$\begin{aligned}
z^{ab} &\sim N(O_k; I_K) \\
z^a, z^b &\sim N(O_k; I_K) \\
y^a &\sim N(\Lambda^a z^{ab} + B^a z^a; \Psi^a) \\
y^b &\sim N(\Lambda^b z^{ab} + B^b z^b; \Psi^b)
\end{aligned} \quad (1)$$

Where $B^j \in \mathbb{R}^{p^j \times k}$, $\Lambda^j \in \mathbb{R}^{p^j \times k}$ and $\Psi^j \in \mathbb{R}^{p^j \times p^j}$. This can be reformulated as a factor analysis problem,[16] thus, suggesting that inference in the PCCA can be performed using expectation-maximization (EM), where the parameters are updated using the following tilling as seen in **Equation 2**:

$$\Lambda^* = \sum_i \left(y_i \mathbb{E}_{z|y_i}[z|y_i]^T\right) \left(\mathbb{E}_{z|y_i}[zz^T|y_i]\right)^{-1}$$

$$\Psi^* = \sum_i \frac{1}{n} diag\left(y_i y_i^T - \Lambda^* \mathbb{E}_{z|y_i}[z|y_i] y_i^T\right) \quad (2)$$

Once the shared and view-specific latent variables $z = [z^{ab}\ z^a\ z^b]^T$ are derived, the next step is to use the reparameterization trick to sample from the PCCA representation $\hat{y}^j \sim \mathcal{N}(\Lambda^{j^*} z^{ab} + B^{j^*} z^j; \Psi^{j^*})$ and obtain embedding samples $\hat{y}^j$. This step ensures that the Monte Carlo estimate of the expectation is distinct with respect to the encoder parameters and, thus, the model can be trained in an end-to-end fashion by defining the following loss function in **Equation 3**:

$$\mathcal{L} = \frac{1}{n} \sum_{i=1}^{n} \left(\|\hat{x}_i^a - x_i^a\|_2^2 + \|\hat{x}_i^b - x_i^b\|_2^2\right) + \gamma(\|\Lambda^b\|_1 + \|\Lambda^{ab}\|_1) \quad (3)$$

In the formulation described in this section, there are five hyperparameters ($p^a, p^b, k^{ab}, k^a$ and $k^b$) determining the dimensions of the modality embeddings and latent space. In this case: $y \in \mathbb{R}^p$ such that $p = p^a + p^b$, the latent space $z \in \mathbb{R}^k$ where $k = k^{ab} + k^a + k^b$, $\Lambda \in \mathbb{R}^{p \times k}$ and $\Psi \in \mathbb{R}^{p \times p}$. To identify the best set of hyperparameters we did a grid search $p \in \{5, 10, 25, 50\}$ and k$\in \{2,3,5,10\}$ such that $k \leq p$ was always satisfied and we selected the smallest number that resulted in a high image and metabolomics reconstruction. This was found to be $p^a = p^b = 10$ and $k^{ab} = k^a = k^b = 3$, such that $p = 20$ and $k = 9$, through the loss function defined in Equation 3 and the reconstruction of both modalities as seen in Figure 4.

## 4.4 Analysis of dataset for testing image embeddings for histology classification and survival

For this section, we trained our model on the TCIA cohort of (n=203) and then performed the test on the test section of TCIA and three different datasets from the (n=320) Royal Marsden Hospital(RMH), (n=128) Guy's and St Thomas' Hospital (GSTT) and (n=101) Imperial College Healthcare Trust (ICHT).[23] We first filtered the datasets only to have patients with AC and SCC histology.

As a baseline for feature quality, we used TextLab 2.0 software to extract 665 features from the lesion. The methods in **Table 6** were applied to features extracted using the DPCCA model and TextLab 2.0, the latter for radiomics analysis.

**Table 6.** Summary of the feature selection and prediction methods used, if (*) method is used for classification, (**) otherwise method is used for both classification and survival.

| Acronym | Feature selection methods | Acronym | Prediction methods |
|---|---|---|---|
| VAR* | Variance | GNB* | Gaussian naïve baye |
| RELF* | Relief | MNB* | Multinomilive bayes |
| MI* | Mutual information | BNB* | Bernoulli naïve bayes |
| mRMR* | Minimum redundancy maximum relevance ensemble | KNN* | K-nearest neighbourhood |
| ETE* | Extra tree ensemble | RF* | Random forest |
| GBDT* | Gradient boosting decision tree | BAG* | Bagging |
| TSQ* | T-test score | DT* | Decision tree |
| CHSQ* | Chi-square score | GBDT* | Gradient boosting decision tree |
| EN* | Elastic net | Adaboost* | Adaptive boosting |
| LASSO | Least absolute shrinkage and selection operator | XGB* | Xgboost |
| WLCX* | Wilcoxon | LDA | Linear discriminant analysis |
| L$^1$-SVM* | L$^1$- based linear support vector machine | LGR | Logistic regression |
| L$^1$-LGR* | L$^1$ -based logistic regression | Linear-SVM | Linear support vector machine |
| JMIM** | Joint mutual information maximisation | RBF-SVM | Radial basis function support vector machine |
| RFVH** | Random forest with variable hunting | MLPC | Multi-layer perceptron |
| RFVH - IMP** | Random forest with variable hunting and Gini impurity corrected variable importance | BGLM_CoxPH | Boosting gradient linear models |
| RF | Random forest variable hunting with maximal depth | BGLM_Cindex | Boosting gradient linear models |
| Spearman** | Spearman correlation | BGLM_Weibull | Boosting gradient linear models |
| Person** | Pearson correlation | BT_CoxPH | Boosting trees |
| Kendall** | Kendall rank correlation | BT_Weibull | Boosting trees |
| Random** | Random (null hypothesis) | Cox_Lasso | Cox lasso |
| | | Cox_Net | Cox net |

| | | |
|---|---|---|
| | Cox | Cox proportional hazard |
| | RSF | Random survival forest |
| | MSR_RF | Random forest using maximally selected rank statistics |
| | ET_RF | Random forest with extra trees |

There exists a wide range of feature selection and machine learning techniques. Identifying the feature selection and machine learning algorithm is task-dependent and critical when developing clinically applicable models. Therefore, we combined different feature reduction techniques for the classification and survival tasks.[33] To find the best combination, we performed a ten-fold cross-validation using the training split of the TCIA data. Then, we used the average accuracy to select the best feature reduction and machine learning algorithms. The acronyms of each feature section, classification and survival method are defined in Table 6.

For the histology classification task on radiomic features, we selected 15 feature selection methods and combined them with 12 machine-learning classifiers based on previous related research.[33, 34] The filter selection methods consisted of univariate and multivariate filter methods, which are classifier-independent and embedded methods such as penalty and tree-based methods, which incorporated the feature selection in the training process. For the classification task, we selected a broad range of methods as suggested by previous studies. We used a cross-combination strategy to select the method with the best mean F1 score across the ten-fold validation.[35] The feature selection and classification task were performed using the scikit-learn package in Python.[36]

In the survival analysis, we used a different set of feature selection methods and models as suggested by the literature and included the Cox proportional hazard model as a benchmark.[37] These specifics are capable of handling censored, heterogenous and high-dimensional data. The machine learning algorithms selected in this section can be divided into four categories: penalized Cox regression, boosted Cox regression (GLM), boosted based on trees and random forests. To select the best combination, we used a cross combination strategy to select the method with the best mean C-index across the ten-fold validation. The feature selection and classification task were performed using the 'survival' package in R. [38]

The best feature selection and machine learning model combination were then selected to perform histology subtype classification and survival analysis using the TCIA validation cohort and the three external datasets, as reported in Tables 3 and 4.

**Acknowledgements**

The authors would like to thank the following organizations for providing data for the work reported: University Hospital Reina Sofia, Spain (metabolomics and CT scans), The Cancer Imaging Archive, USA (CT scans), and OCTAPUS-AI (Optimizing Cancer Treatment Surveillance and Ascertaining cause of Pneumonitis Using Artificial Intelligence).

This article is independent research funded by the Medical Research Council and Imperial STRATiGRAD PhD programme. The authors acknowledge support from Medical Research Council Grant (MR/M015858/1), Imperial College NIHR Biomedical Research Centre award (WSCC_P62585), Imperial College Experimental Cancer Medicines award (C1312/A25149) and National Cancer Imaging Translational Accelerator (C2536/A28680). This work was funded by the Spanish Ministerio de Ciencia e Innovación (MICINN, PID2021-124314OB-I00), and Junta de Andalucía-Consejería de Conocimiento, Investigación y Universidad (P20_00470) grants to MAC. The funding source did not have a role in the study design, data collection, analysis, or interpretation of data.

OCTAPUS-AI Authors:
Thomas G. Charlton[10], Benjamin Hunter[2,8,9], Charleen Chan[8], Merina Ahmed[11], Matthew Orton[12], Jason Lunn[8], Simon J Doran[8], Shahreen Ahmad[10], Fiona McDonald[2,8], Imogen Locke[11], Danielle Power[13], Matthew Blackledge[14]

[8]Institute of Cancer Research NIHR Biomedical Research Centre, London, UK. [9]Cancer Imaging Centre, Department of Surgery & Cancer, Imperial College London, Du Cane Road, London, W12 0NN, UK.[10]Guy's Cancer Centre, Guy's and St Thomas' NHS Foundation Trust, Great Maze Pond, London, SE19RT, UK.[11]Lung Unit, The Royal Marsden NHS Foundation Trust, Downs Road, Sutton, SM25PT, UK. [12]Artificial Intelligence Imaging Hub, Royal Marsden NHS Foundation Trust, Downs Road, Sutton, SM25PT, UK. [13]Department of Clinical Oncology, Charing Cross Hospital, Fulham Palace Road, London, W6 8RF, UK. [14]Radiotherapy and Imaging, Institute of Cancer Research, 123 Old Brompton Road, London, SW7 3RP, UK

**Data availability**
Original metabolomics and CT-scan data are not publicly available but can be accessed via a request to the corresponding authors and/or OCTAPUS-AI.

**Citations**

Intro
[1] H. LU, M. Arshad, A. Thornton, G. Avesani, P. Cunnea, E. Curry, F. Kanavati, J. Liang, K. Nixon, S. T. Williams, M. A. Hassan, D. D. L. Bowtell, H. Gabra, C. Fotopoulu, A. Rockall & E.O. Aboagye, *Nature communications*. **2019**, 10.1: 76.

[2] B. Hunter, M. Chen, P. Ratnakumar, E. Alemu, A. Logan, K. Linton-Reid, D.Tong, N. Senthivel, A. Bhamani, S. Bloch, S. V. Kemp, L. Boddy, S. Jain, S. Gareeboo, B. Rawal, S. Doran, N. Navani, A. Nair, C. Bunce, S. Kaye, M. Blackledge, E. O. Aboagye, A. Devaraj & R. W. Lee, *EBioMedicine*.**2023**, *86*, 104344.

[3] K. Bera, N. Braman, A. Gupta, V. Velcheti, A. Madabhushi, *Nature Reviews Clinical Oncology*. **2022**, 19.2: 132-146.

[4] K. Mao, R. Tang, X. Wang, W. Zhang, H. Wu, *Complexity*. **2018**, 3078374.

[5] T. Kadir, F. Gleeson, *Translational lung cancer research*. **2018**, 7.3: 304.

[6] D. Ardila, A. P. Kiraly, S. Bharadwaj, B. Choi, J. J. Reicher, L. Peng, D. Tse, M. Etemadi, W. Ye, G. Corrado, D. P. Naidich & S. Shetty, *Nature medicine*. **2019**, 25.6: 954-961.

[7] D. Kumar, A. Wong, D. A. Clausi, *Conference on Computer and Robot Vision*. **2015**, 12, 133:138.

[8] M.M. Boubnovski, M. Chen, K. Linton-Reid, J.M. Posma, S.J. Copley, E.O. Aboagye, Clinical Radiology. **2022**, 77.8: e620-e627.

[9] I. Fornacon-Wood, C. Faivre-Finn, J. P. O'Connor, G. J. Price, *Lung Cancer*. **2020**, 146, 197:208.

[10] A. Ibrahim, S. Primakov, M. Beuque, H.C. Woodruff, I. Halilaj, G. Wu, T. Refaee, R. Granzier, Y. Widaatalla, R. Hustinx, F.M. Mottaghy, P. Lambin, *Methods*. **2021**, *188*, 20-29.

[11] A. N. Frix, F. Cousin, T, Refaee, F. Bottari, A. Vaidyanathan, C. Desir, W. Vos, S. Walsh, M. Occhipinti, P. Lovinfosse, R.T.H. Leijenaar, R. Hustinx, P. Meunier, R. Louis, P. Lambin, J. Guiot, *Journal of Personalized Medicine*, **2021**, 11.7: 602.

[12] I. Shiri, M. Amini, M. Nazari, G. Hajianfar, A. H. Avval, H. Abdollahi, M. Oveisi, H. Arabi, A. Rahmim, H. Zaidi, *Computers in biology and medicine*, **2022**, 142: 105230.

[13] F. Carrillo-Perez, J.C. Morales, D. Castillo-Secilla, O. Gevaert, I. Rojas, L.J. Herrera, *Journal of Personalized Medicine.* **2022***,* 12.4, 601.

[14] S. Takahashi, K. Asada, K. Takasawa, R. Shimoyama, A. Sakai, A. Bolatkan, N. Shinkai, K. Kobayashi, M. Komatsu, S. Kaneko, J. Sese, R. Hamamoto, *Biomolecules*. **2020** 10.10, 1460.

[15] T. Y. Lee, K.Y. Huang, C.H. Chuang, C.Y. Lee, T.H. Chang, *Computational Biology and Chemistry*. **2020**, *87*:107277.

[16] G. Gundersen, B. Dumitrascu, J. T. Ash, B. E. Engelhardt, In *Proceedings of The 35th Uncertainty in Artificial Intelligence Conference*. **2020**, 35, 945:955.

[17] P. Moreno, C. Jimenez-Jimenez, M. Garrido-Rodrıguez, M. Calderon-Santiago, S. Molina, M. Lara-Chica, F. Priego-Capote, A. Salvatierra , E. Munoz, M. A. Calzado, *Molecular oncology.* **2018**, 12, 1778:1796.

[18] Y. Li, X. Wu, P. Yang, G. Jiang, Y. Luo, *Genomics, Proteomics & Bioinformatics*, **2022**, 20.5 , 850-866.

[19] A. Kumar, B. B. Misra. *Proteomics.* **2019***,* 19.21-22:1900042.

Results

[20] S. G.  Armato III, C. R. Meyer, M. F. McNitt-Gray, G. McLennan, A. P. Reeves, B. Y. Croft, L. P. Clarke, The RIDER Research Group, Clinical Pharmacology & Therapeutics. 2008, 84.4: 448-456.

[21] B. Zhao, L. P. James, C. S. Moskowitz, P. Guo, M. S. Ginsberg, R. A. Lefkowitz, Y. Qin, G. J. Riely, M. G. Kris, L. H. Schwartz,  Radiology. 2009, 252.1: 263-272.

[22]  H. J. W. L. Aerts, E. R. Velazquez, R. T. H. Leijenaar, C. Parmar, P. Grossmann, S. Carvalho, J. Bussink, R. Monshouwer, B. Haibe-Kains, D. Rietveld, F. Hoebers, M. M. Rietbergen, C. R. Leemans, A. Dekker, J. Quackenbush, R. J. Gillies, Philippe Lambin, *Nature communications*. **2014**, 5.1: 4006.

[23] S. Hindocha, T. G. Charlton, K. Linton-Reid, B. Hunter, C. Chan, M. Ahmed, Emily J. Greenlay, M. Orton, C. Bunce, J. Lunn, S. J. Doran, S. Ahmad, F. McDonald, I. Locke, D. Power, M. Blackledge, R. W. Lee, E. O. Aboagye, *NPJ Precision Oncology*. **2022**, 6.1: 77.

[24] P. Langley, Philip, E. J. Bowers, A. Murray, *IEEE transactions on biomedical engineering*. **2009**, 57.4, 821-829.

[25] A. Rosato, L. Tenori, M. Cascante, P.R.C. Atauri, V.A.M. Santos, E. Saccenti, *Metabolomics*. **2018**, *14*, 1-20.

[26] J. E. van Timmeren, D. Cester, H. Alkadhi, B. Baessler. *Insights into Imaging*. **2020**. *11*(1), 1-16.

[27] X. Teng, J. Zhang, Z. Ma, Y. Zhang, S. Lam, W. Li, H. Xiao, T. Li, B. Li, T. Zhou, G. Ren, F. K. Lee, K. Au, V. H. Lee, A. T. Chang, J. Cai, *Frontiers in Oncology*. **2022**. *12*.

[28] D. Leng, L. Zheng, Y. Wen, Y. Zhang, L. Wu, J. Wang, M. Wang, Z. Zhang, S. He, X. Bo *Genome Biology*. **2022**, *23*(1), 1-32.

[29] B. Hunter, M. Chen, P. Ratnakumar, E. Alemu, A. Logan, K. Linton-Reid, D. Tong, N. Senthivel, A. Bhamani, S. Bloch, S. V. Kemp, L. Boddy, S. Jain, S. Gareeboo, B. Rawal, S. Doran, N. Navani, A. Nair, C. Bunce, S. Kaye, M. Blackledge, E. O. Aboagye, A. Devaraj, R. W. Lee, *EBioMedicine*. **2022**, 86: 104344.

[30] F. Pérez-García, R. Sparks, S. Ourselin, *Computer Methods and Programs in Biomedicine*. **2021**, 208: 106236.

[31] A. Hosny, C. Parmar, T. P. Coroller, P. Grossmann, R. Zeleznik, A. Kumar, J. Bussink, R. J. Gillies, R. H. Mak, H. J. W. L. Aerts, *PLoS medicine*. **2018**, 15.11: e1002711.

[32] D. Nie, R. Trullo, J. Lian, L. Wang, C. Petitjean, S. Ruan, Q. Wang; D.9 Shen, *IEEE Transactions on Biomedical Engineering*, **2018**, 65.12: 2720-2730.

[33] P. Sun, D. Wang, V. C. Mok, L. Shi, *IEEE Access*. **2019**, 7: 102010-102020.

[34] P. Yin, N. Mao, C. Zhao, J. Wu, C. Sun, L. Chen, N. Hong, *European radiology*. **2019**, *29*, 1841-1847.

[35] M. Destito, A. Marzullo, R. Leone, P. Zaffino, S. Steffanoni, F. Erbella, F. Calimeri, N. Anzalone, E. De Momi, A.J.M Ferreri, T. Calimeri, M.F. Spadea, *Bioengineering*, **2023**, 10.3: 285.

[36] F. Pedregosa, G. Varoquaux, A. Gramfort, V. Michel, B. Thirion, O. Grisel, M. Blondel, P. Prettenhofer, R. Weiss, and V. Dubourg, *J. Mach. Learn*. **2011**, 12, 2825–2830.

[37] A. Spooner, E. Chen, A. Sowmya, P. Sachdev, N. A. Kochan, J. Trollor, H. Brodaty, *Sci. Rep*. **2020**, 10, 20410.

[38] T. M. Therneau, T. Lumley, *R Top Doc*. **2015** 128.10, 28-33.